\documentclass[manuscript]{aastex}

\shorttitle{Data characterization using artificial-star tests} 
\shortauthors{Hu et al.}  

\title{Data characterization using artificial-star tests: performance
evaluation}

\begin{document}

\author{Yi Hu,\altaffilmark{1,2} Licai Deng,\altaffilmark{1} Richard
de Grijs,\altaffilmark{3} and Qiang Liu\altaffilmark{1}}

\email{huyi@bao.ac.cn}

\altaffiltext{1}{National Astronomical Observatories, Chinese Academy of
Sciences, Beijing 100012, P. R. China}
\altaffiltext{2}{Graduate University of the Chinese Academy of Sciences, Beijing
100012, P. R. China}
\altaffiltext{3}{Kavli Institute for Astronomy and Astrophysics,
Peking University, Beijing 100871, P. R. China}

\begin{abstract}
Traditional artificial-star tests are widely applied to photometry in
crowded stellar fields. However, to obtain reliable binary fractions
(and their uncertainties) of remote, dense, and rich star clusters,
one needs to recover huge numbers of artificial stars. Hence, this
will consume much computation time for data reduction of the images to
which the artificial stars must be added. In this paper, we present a
new method applicable to data sets characterized by stable,
well-defined point-spread functions, in which we add artificial stars
to the retrieved-data catalog instead of the raw images. Taking the
young Large Magellanic Cloud cluster NGC 1818 as an example, we
compare results from both methods and show that they are equivalent,
while our new method saves significant computational time.
\end{abstract}

\keywords{Star Clusters and Associations -- Data Analysis and Techniques}

\section{Introduction}

The artificial-star-test technique has been widely applied to
photometry of crowded stellar fields over the last few decades. By
adding artificial stars to the original, raw images and subsequently
recovering them, such tests have proved an efficient means of counting
stars as a function of magnitude and measuring the associated
photometric accuracies (e.g., Bolte 1989, 1994; Bergbusch 1996;
Sandquist et al. 1996; Da Rocha et al. 2002; Hargis et al. 2004;
Fekadu et al. 2007). Modern computers (`central processing units:'
CPUs) are becoming ever more powerful, thus providing opportunities to
perform artificial-star experiments to characterize the
color-magnitude-diagram (CMD) morphologies of remote, rich star
clusters and obtain cluster binary fractions by comparing CMDs based
on artificial and real stars (e.g., Rubenstein \& Bailyn 1997; Zhao \&
Bailyn 2005; Hu et al. 2010).

However, even with the current generation of CPUs, this method still
consumes significant computation time if one wants to use it to derive
cluster binary fractions, for which many more (artificial) stars are
needed than to obtain observational completeness levels. For instance,
Bolte (1994) used $5.5 \times 10^4$ artificial stars (in 450 test
runs, i.e., on the order of 100 stars per trial, which must be
compared to 200 artificial stars per run used in this paper; see
below) to calculate the completeness fraction of his observations of
the globular cluster M30, where they recovered 5674 real stars within
the range $13 \le V \le 22$ mag. Similarly, Bellazzini et al. (2002)
required $1.5 \times 10^6$ artificial stars ($17 \le V \le 25$ mag) to
obtain an estimate of the binary fraction in NGC 288 (compared with
5766 observed stars in the cluster's central regions and 2013 near its
half-light radius, with $13 \le V \le 25.5$ mag).

With such large artificial-star catalogs,\footnote{By `catalog' we
mean the ensemble of data resulting from object-recovery routines
applied to crowded-field imaging data, including stellar positions,
magnitudes, and the associated uncertainties.} both the addition of
stars to the images and the subsequent data reduction required for
recovery of these artificial stars will be enormously
time-consuming. As an altenative, Hu et al. (2010) developed and
applied a novel, efficient method (in which they added artificial
stars to the catalog instead of the images) to obtain the binary
fraction of the young Large Magellanic Cloud cluster NGC 1818.

In this paper, we use the raw images of NGC 1818, taken with the Wide
Field and Planetary Camera-2 (WFPC2) on board the {\sl Hubble Space
Telescope (HST)} as an example to benchmark and compare the
performance of both artificial-star-test methods. The photometric data
are discussed in \S 2. The commonly used method of adding artificial
stars to images and our newly developed method to correct for stellar
blends and superpositions based on catalog handling are presented in
\S 3. A detailed comparison of the two methods and validation of our
new approach are provided in \S 4.

\section{Data reduction}

Our example data set was obtained from {\sl HST} program GO-7307 (PI
Gilmore), which included three images in both the F555W and F814W
filters (with exposure times of 800, 800, and 900 s for each filter),
centered at the cluster's half-light radius. The observations were
reduced as described in Hu et al. (2010) using {\sc HSTPhot} (version
1.1, May 2003; Dolphin 2005).

In the left-hand panel of Fig. 1 we show the resulting
CMD, reduced using {\sc HSTPhot} with the point-spread-function
(PSF)-fitting option. The corresponding CMDs, reduced using both {\sc
HSTPhot} with the aperture-photometry (`APP') option and {\sc iraf}'s
{\sc apphot} task (Liu et al. 2009), are shown in the middle and
right-hand panels. Within the magnitude range from approximately 19 to
27 in both the F555W and F814W filters (an unambiguous detection in
both filters was required), we recovered a total of 8756 and 7773
stars using PSF fitting and aperture photometry based on {\sc
HSTPhot}, respectively, while Liu et al. (2009) recovered 7332 stars
from the same data set (but for $V \le 26$ mag). Our new CMDs are much
cleaner (we rejected stars with $\chi>1.5$, which is a measure of the
signal-to-noise ratio and a level recommended in the {\sc HSTphot}
manual; cf. Dolphin 2000) than that of Liu et al. (2009), who did not
apply a cut in signal-to-noise ratio, and the cluster's main sequence
is much tighter. Fig. 3 shows the standard deviations of
the photometric uncertainties as a function of stellar magnitude. It
is clear that, for the same magnitude range, the uncertainties
associated with photometry of stars reduced by {\sc HSTPhot} are
smaller than the equivalent values resulting from application of {\sc
iraf/apphot}. This indicates that our most recently obtained
photometry is more precise than that published by Liu et al. (2009),
because the {\sc HSTphot} software package is much better at properly
handling stellar photometry in crowded {\sl HST} fields than {\sc
iraf}'s {\sc apphot} routine.

Note that although the PSF-fitting approach recovered approximately a
thousand more stars than the number detected by aperture photometry,
the number of stars brighter than $m_{\rm F555W} = 25$ mag (but recall
that stars must be detected in both the F555W and F814 filters to be
included) was 5974 versus 5791 for the PSF-fitting and
aperture-photometry approaches, respectively. The corresponding growth
curves for all three reduction methods are shown in
Fig. 2 and tabulated in Table 1. The
numbers of stars resulting from PSF fitting and aperture photometry
are comparable; they are much more numerous than those resulting from
the use of {\sc iraf/apphot} (Liu et al. 2009). Therefore, we conclude
that {\sc HSTPhot} is more appropriate than standard {\sc iraf} tasks
to deal with stellar photometry in crowded {\sl HST} images and that
in crowded fields, such as for NGC 1818, there is no significant
difference in performance between the two photometric methods
supported by {\sc HSTPhot} (except for faint stars close to the
detection limit; see also \S 4).

\section{Artificial-star tests}

We developed our own routines, which we used instead of the {\sc
iraf/addstar} task, to add artificial stars to the images. The masses
of the artificial stars were selected from a Salpeter (1955)-like mass
function, which was shown to be appropriate for this cluster by de
Grijs et al. (2002), Liu et al. (2009), and Hu et al. (2010) for
stellar masses in excess of 0.5 to $1 M_\odot$. For each star, we
calculated the magnitude and color by interpolation of the
best-fitting Girardi et al. (2000) isochrone for a cluster age of 25
Myr and a metallicity of $Z=0.008$ (where $Z_\odot = 0.019$). Next, we
converted the stellar absolute magnitude in the $V$ band ($M_V$) to a
total pixel value ({\it PV}) using $M_V = -2.5\log \mbox{\it PV} +
M_0$, where $M_0$ is the zero-point offset (cf. Hu et al. 2010): $M_0
= -2.5 \log({\rm PHOTFLAM}) + {\rm PHOTZPT}$, where PHOTFLAM and
PHOTZPT ($= -21.1$) are image-header keywords. Since the CCD's
charge-transfer efficiency (CTE) is a function of position, we used
the prescription of Whitmore et al. (1999) to calculate CTE
corrections. We calculated scaled {\sl HST}/WFPC2 PSFs using {\sc
TinyTim} (Krist \& Hook 2004) for different filters, positions, and
stellar types, before adding appropriately constructed artificial
stars to the science images. To do so, we used the {\sc TinyTim} PSFs
to expand the point sources to extended sources ($64\times 64$
pixels$^2$ for the Planetary Camera chip, PC1, and $30\times 30$
pixels$^2$ for the Wide Field chips, WF1, 2, and 3). The artificial
stars were uniformly spread across the images. To avoid interference
between artificial stars, we only added 50 stars at any one time to a
given chip. In addition, we ensured that each artificial star occupied
a unique $64\times 64$ (or, as appropriate, $30\times 30$) pixels$^2$
subfield. Finally, we used the same approach (i.e., identical to that
used for the analysis of our science data) to recover the added
artificial stars from the artificial-star images.

An alternative method to perform artificial-star tests is based on
adding stars to the catalog rather than the images (Hu et
al. 2010). By applying this method, we do not need to use PSFs to add
artificial stars repeatedly to the images and reduce hundreds of
thousands of artificial-star images. This will, therefore, potentially
save much computation time. Unlike while adding artificial stars to
the images, the effects of superposition are automatically
included. When we add artificial stars to the spatial-distribution
diagram obtained from the catalog, we assume that if an artificial
star is located at a distance from any real star of less than 2 pixels
(corresponding to the size of our aperture), it is `blended.' In this
case, we simply add the fluxes of the artificial and real stars to
recalculate the artificial star's magnitude and color. If the output
artificial star is 0.752 mag (or more) brighter than the input value,
it is assumed to be blended and rejected from the output
artificial-star catalog.

\section{Comparison}

We performed extensive artificial-star experiments to test the
validity of our assumption that stellar images centered within 2
pixels of each other cannot be distinguished individually. To do so,
we added sets of two artificial stars to a blank WF2 subfield (i.e., a
subfield without any real stars detectable above the noise level, and
with realistic noise characteristics). We adopted separations between
both stars of $\Delta d = 0, 1, 2, 3, 4, 5, 10, 15$, and 20
pixels. The resulting images, for a range of stellar-mass ratios, are
shown in Figs 4, 5, and 6. We then performed photometry using {\sc HSTphot} and
attempted to recover the fluxes of both stars. Fig. 7
shows the magnitudes of the recovered artificial stars as a function
of their separation, assuming input masses of $1.6 M_\odot$ for
both. The recovered magnitude decreases rapidly for separations of
less than 2 pixels and reaches a plateau beyond this separation. The
difference between the combined flux at $\Delta d=0$ pixel and the
recovered magnitudes for the individual artificial stars reaches 0.753
mag at $\Delta d=2$ pixels (in the sense that they are fainter for
this $\Delta d$ than when they are blended). This indicates that our
assumption of a 2-pixel minimum separation to distinguish two stars
individually is a good approximation.

\subsection{Completeness}

We also performed completeness tests. Incompleteness of bright stars
is dominated by superpositions in the crowded stellar environment of
our observations, while for faint stars it is dominated by the
instrument's detection limit. We adopted a Gaussian-type photometric
error for each artificial star (see below), which was added to the
spatial-distribution catalog. We also set a 50\% completeness limit at
$m_{\rm F555W} = 25$ mag. If the output artificial star is 0.752 mag
(or more) brighter than the input value, the star is blended; if its
magnitude is fainter than $m_{\rm F555W} = 25$ mag, it cannot be
detected. Fig. 9 shows the resulting completeness
fraction, calculated by adding 10,000 artificial stars to both the
catalog and the images at any one time.

Both incompleteness fractions agree fairly well for stars brighter
than $m_{\rm F555W} = 23$ mag. Therefore, we conclude that our new
method is adequate for studying the binary fractions of stars that are
at least $\sim 2$ mag brighter than the observational completeness
limit. For fainter stars, our new method cannot be applied with
sufficient reliability to estimate completeness fractions. This is so,
because by adding artificial stars to the catalog, the completeness
fraction will be marginally higher than we would have measured from
adding these stars to the images, because we do not take into account
the presence of cosmic rays and bad pixels.  Therefore, although
it offers a quick, first-order indication of a stellar sample's
observational completeness, our method cannot be used to calculate
realistic, quantitative incompleteness fractions of real observational
data.

\subsection{Error analysis}

When adding artificial stars to the catalog, we assumed that the
photometric errors are distributed in a Gaussian manner. We obtain the
relation between stellar magnitude and the corresponding errors from
the catalog. We used an exponential function to fit the relation
(cf. Hu et al. 2010). To test this assumption, we added 500 artificial
stars of the same mass to the raw images and subsequently recovered
them using standard data-reduction techniques. Fig. 10
shows the distribution of the differences between the input and output
F555W magnitudes (converted to $V$). The mean photometric error is
0.017 mag, while the standard deviation of the distribution in
Fig. 10 is 0.02 mag. Thus, our assumption appears
justified.

\section{Discussion and conclusions}

Modern computers are becoming ever faster. When Bolte (1994)
calculated the completeness fraction of his observations of M30 using
artificial-star tests, it took him nearly one month of computation
time on a Sun workstation. Now, such an experiment
will only take a few hours on a personal computer. However, if we use
artificial-star tests to estimate the binary fraction of a rich star
cluster, we need on the order of 100 times more artificial stars ($1
\times 10^6$) for each run compared to the requirements for
(in)completeness tests. On an Intel Xeon E5430
server, the computation time needed for data reduction {\it only}
(i.e., not including the addition of artificial stars) of a single
{\sl HST}/WFPC2 footprint (consisting of four frames of $800 \times
800$ pixels$^2$) is 1.2 min per core, which means that we need at
least one month just for the data reduction if we were to perform
artificial-star tests on the science images. For instance, to run our
simulations for $5 \times 10^6$ stars (cf. Hu et al. 2010), we would
need to generate 25,000 footprints (each footprint would contain four
times 50 artificial stars). In this case, the data-reduction time
needed amounts to 21 days, i.e., $25,000 \times 1.2/(60\times24)$, for
a single test (e.g., for a single value of the expected binary
fraction in the cluster). On the other hand, by adding artificial
stars to the catalog, we can `recover' as many stars as we need, and
the total computation time is much shorter. For example, adding $5
\times 10^6$ stars only requires 2 hours on a personal computer, again
for a single test run.

Based on extensive tests, we conclude that adding artificial stars to
the spatial-distribution diagram provided by the catalog and adopting
a 2-pixel separation threshold for stellar blends provides a good
approximation to reality. Comparison of the (in)completeness fractions
obtained from both methods underscores our conclusion. Estimating
binary fractions of dense star clusters using artificial-star tests
can safely be done by adding stars to the catalog rather than the
images. The new method is much less time-consuming yet equivalent in
performance to the old method.  We note, however, that this method
can only be used for observational data sets characterized by stable,
well-defined PSFs. This limits its application to space-based
observations; the PSFs of ground-based observations depend not only on
instrumental and detector characteristics, but also on time-varying
atmospheric conditions (`seeing').

\acknowledgments 
LC acknowledges the support by National Natural Science Foundation of
China (NSFC) through grant 10973015, and the Ministry of Science and
Technology of China through grant 2007CB815406. RdG acknowledges
partial research support through NSFC grants 11043006 and 11073001.

\clearpage
\begin{table}
\caption{Growth curves: numbers of stars detected as a function of magnitude and recovery method.}
\begin{center}
\begin{tabular}{cccc}
\hline
 Magnitude range ($m_{\rm F555W}$) & {\sc HSTPhot}/Aperture & {\sc HSTPhot}/PSF & {\sc iraf/apphot} \\
 \hline
 $<19$ & 1    & 1    & 0     \\
 $<20$ & 279  & 263  & 52    \\
 $<21$ & 1037 & 1020 & 722  \\
 $<22$ & 1906 & 1875 & 1659 \\
 $<23$ & 3029 & 3013 & 2811 \\
 $<24$ & 4306 & 4310 & 4144 \\
 $<25$ & 5791 & 5974 & 5639 \\
 $\ge 25$ & 2781 & 1982 & 1692 \\
\hline
\end{tabular}
\end{center}
\end{table}

\clearpage
\begin{figure*}
\includegraphics[width=\columnwidth]{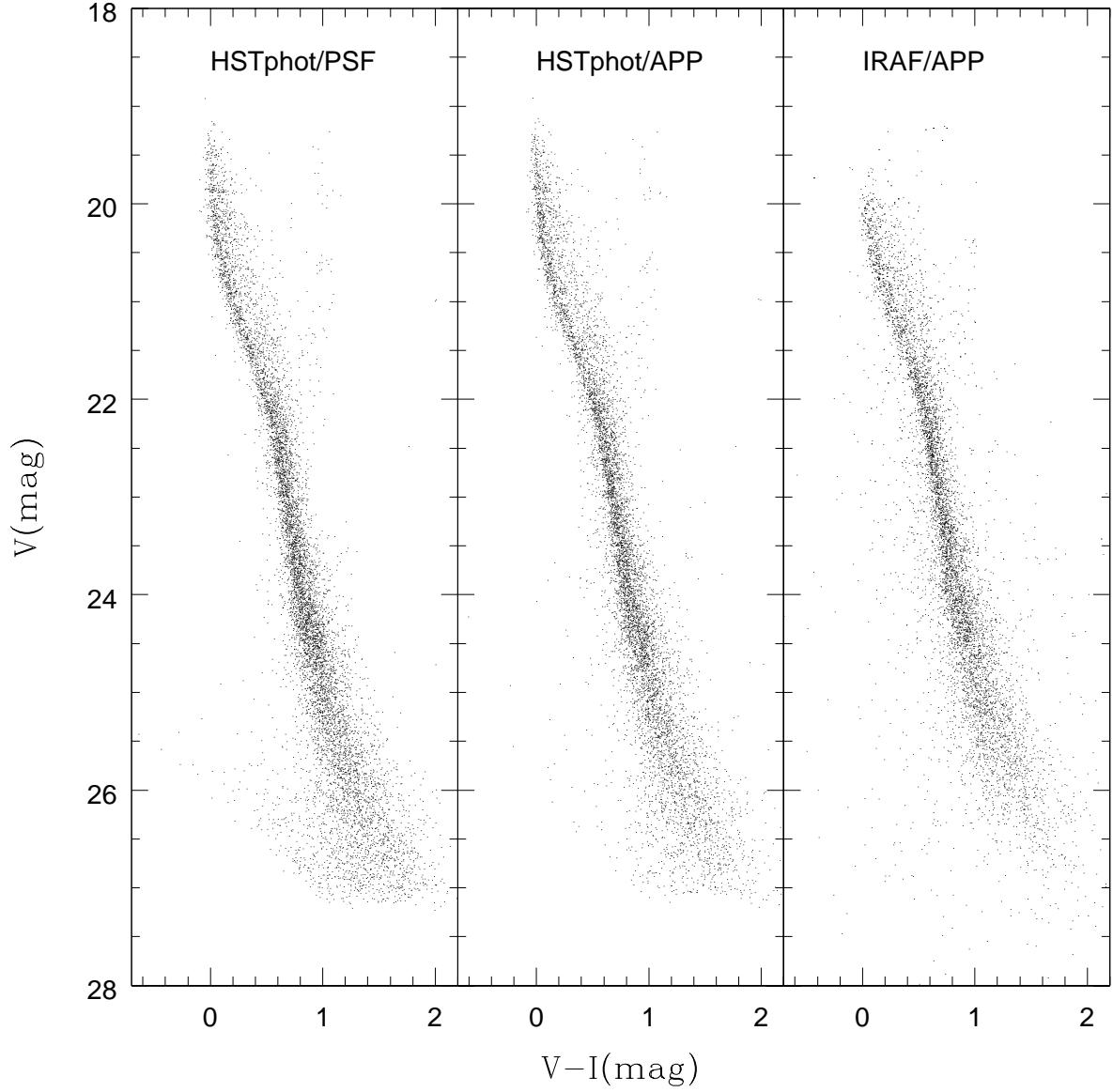}
\caption{NGC 1818 CMDs. (left) {\sc HSTphot}/PSF fitting. (middle)
{\sc HSTphot}/Aperture photometry. (right) {\sc iraf/apphot} from Liu
et al. (2009).}
\end{figure*}

\clearpage
\begin{figure*}
\includegraphics[width=\columnwidth]{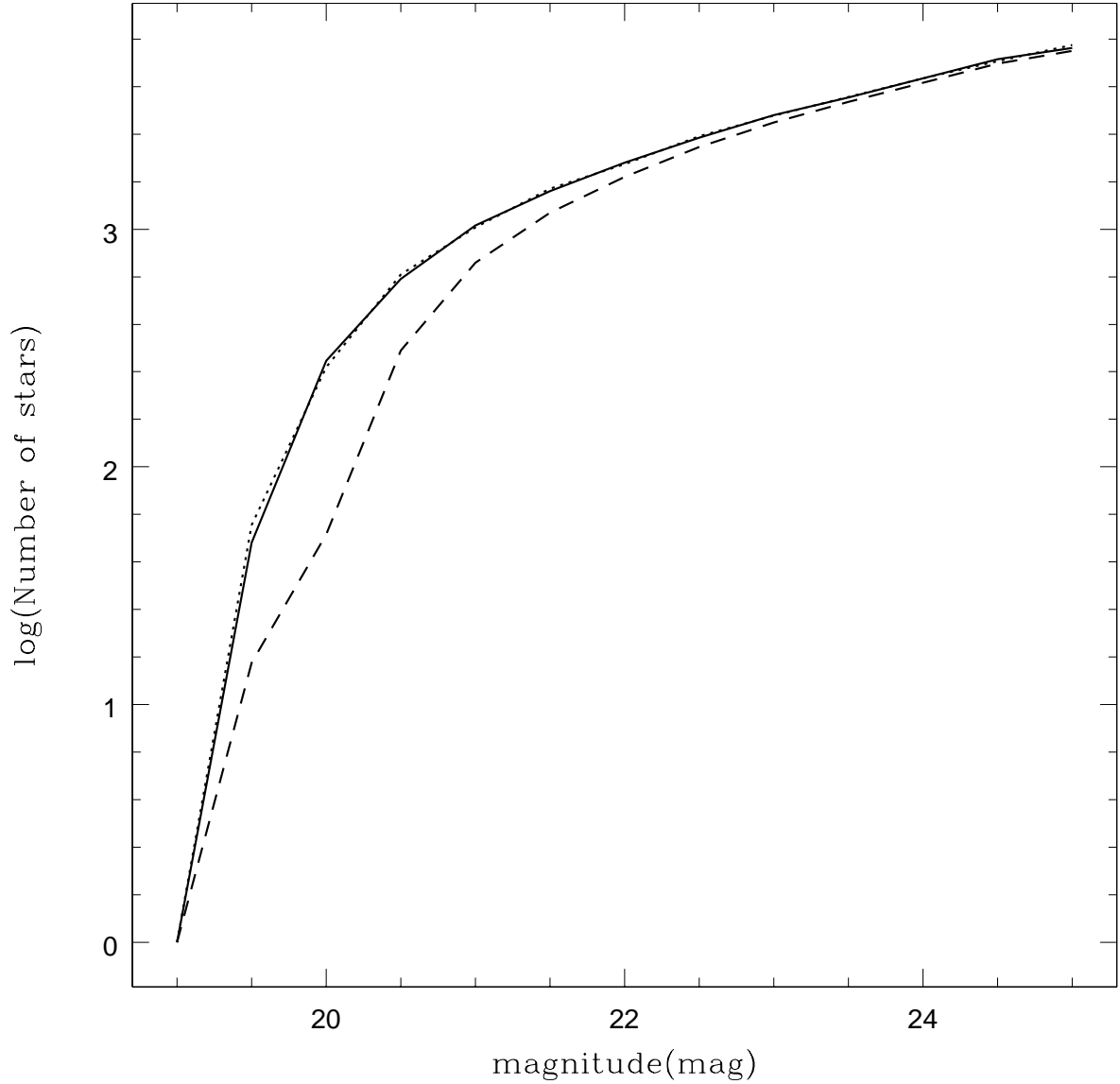}
\caption{Growth curves. The solid, dotted, and short-dashed lines are
for {\sc HSTPhot}/PSF fitting, {\sc HSTPhot}/Aperture photometry, and
{\sc iraf/apphot}, respectively. The solid and dotted lines nearly
overlap, which sugguests that both PSF fitting and aperture photometry
are adequate for analysis of our field. }
\end{figure*}

\clearpage
\begin{figure*}
\includegraphics[width=\columnwidth]{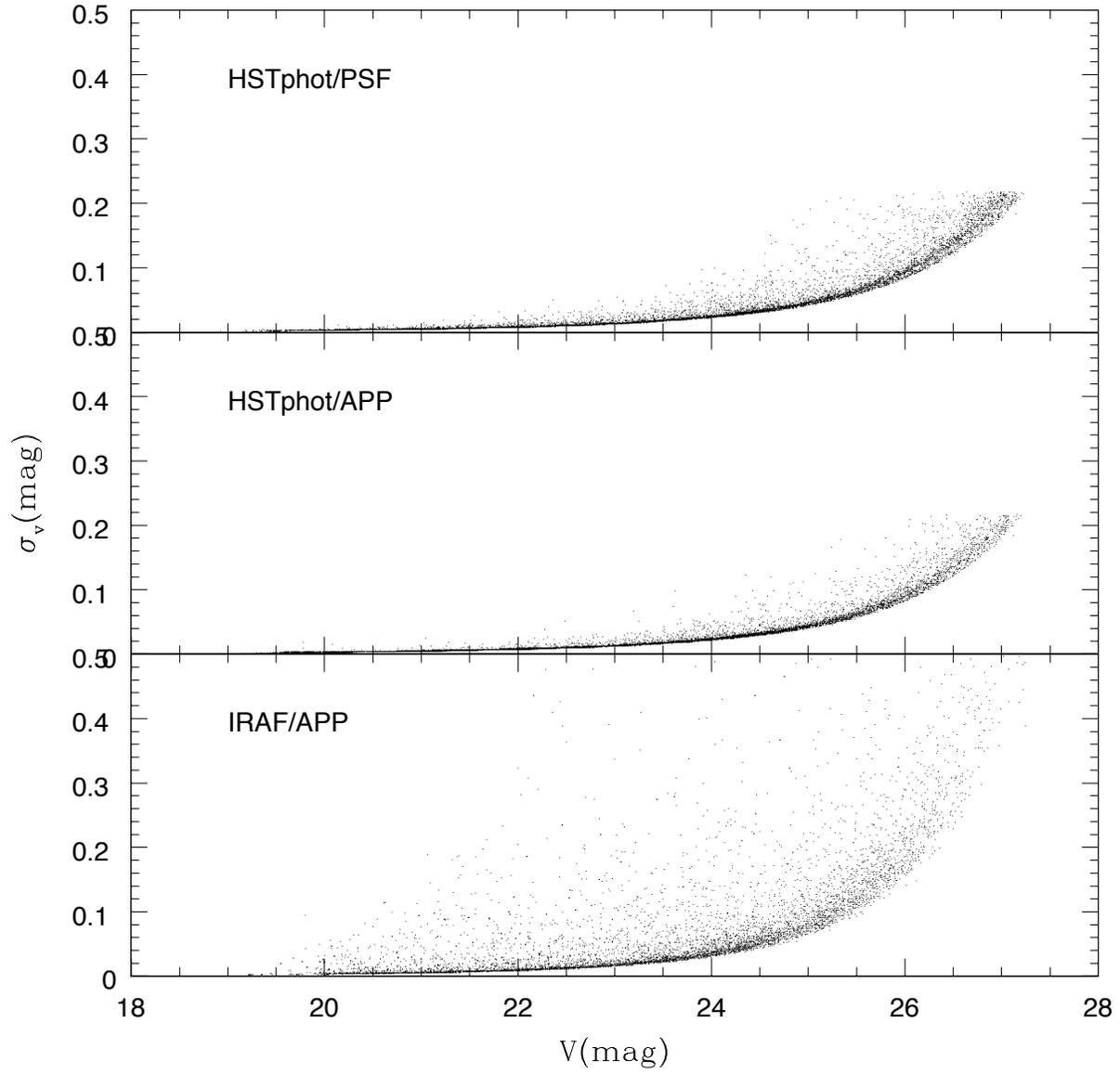}
\caption{Standard deviations of the photometric uncertainties as a
function of stellar magnitude (cf. Hu et al. 2010).}
\end{figure*}

\clearpage
\begin{figure*}
\includegraphics[width=\columnwidth]{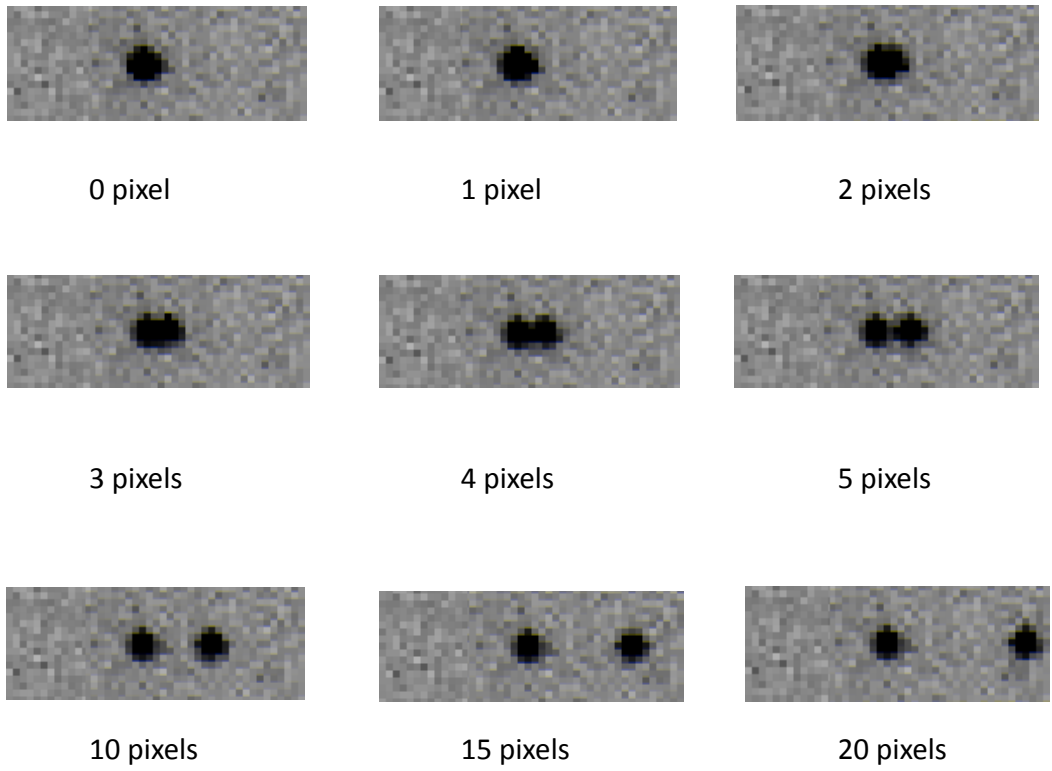}
\caption{Images of artificial stars added to the raw science
frames. The masses of the two stars are both $1.6 M_\odot$, and the
labels refer to the separation between the stars.}
\end{figure*}

\clearpage
\begin{figure*}
\includegraphics[width=\columnwidth]{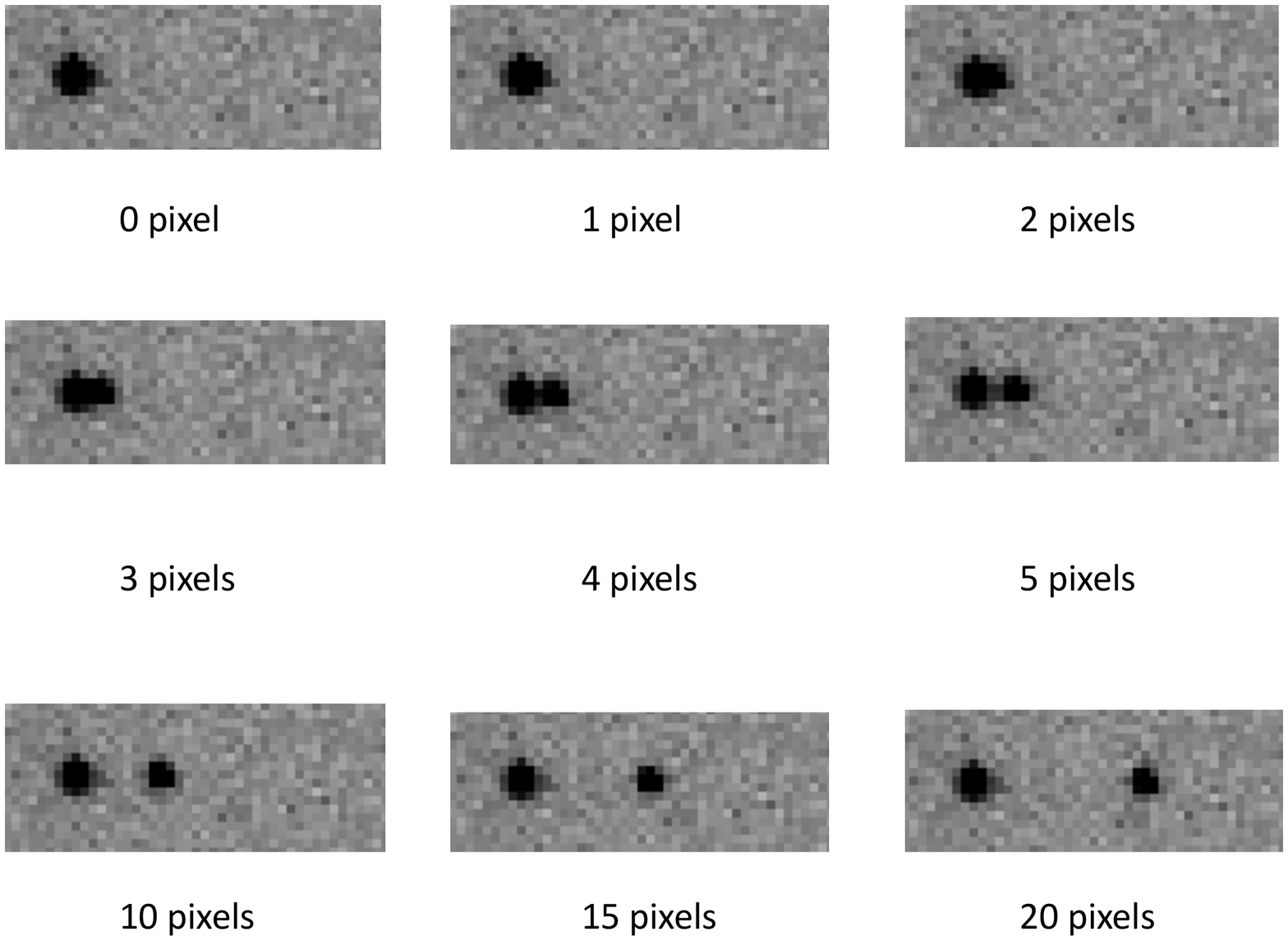}
\caption{Same as Fig. 4. The masses of the two stars are
$1.6$ and $1.3 M_\odot$. }
\end{figure*}

\clearpage
\begin{figure*}
\includegraphics[width=\columnwidth]{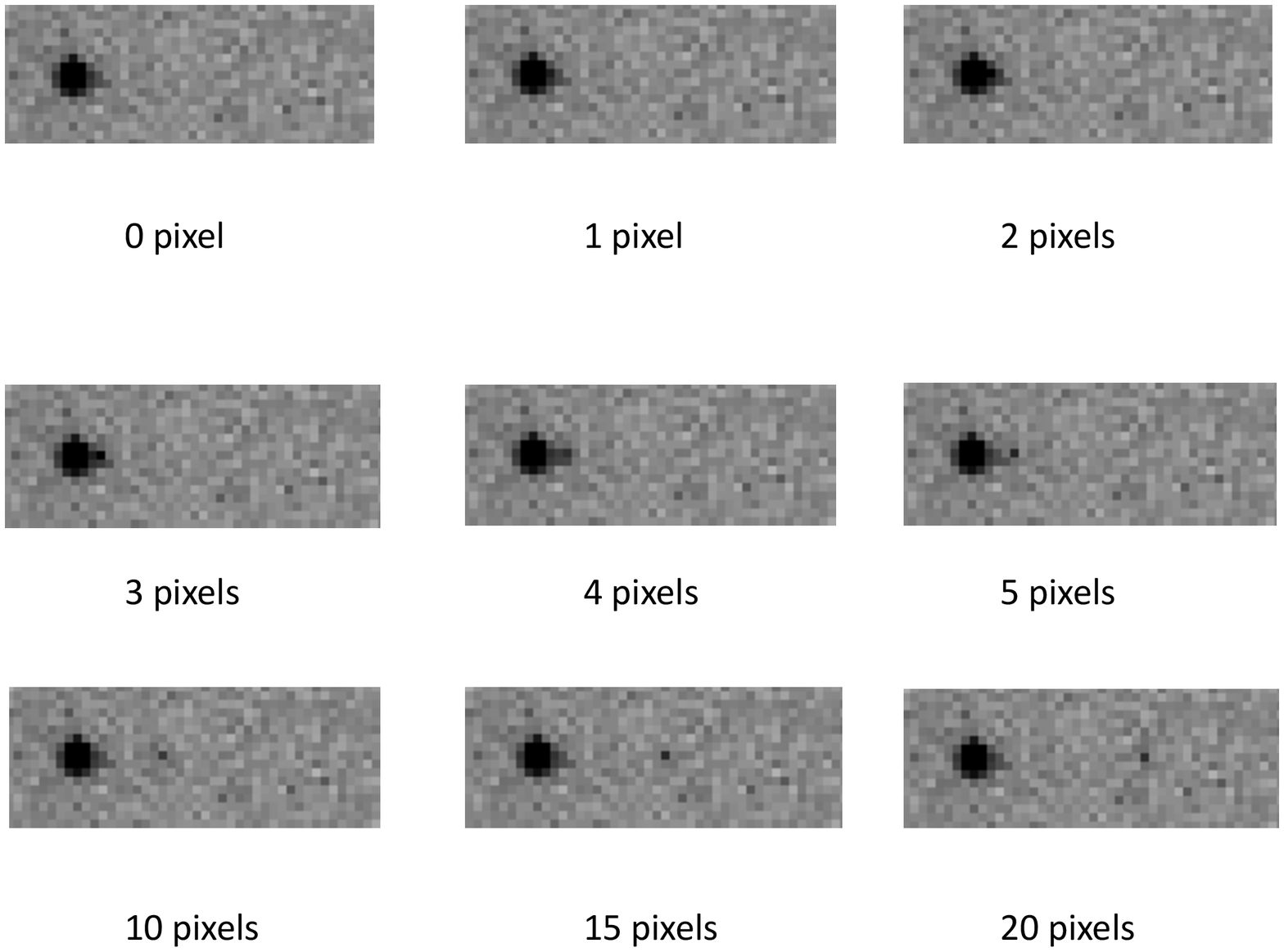}
\caption{Same as Fig. 4. The masses of the two stars are
$1.6$ and $0.8 M_\odot$. }
\end{figure*}

\clearpage
\begin{figure*}
\includegraphics[width=\columnwidth]{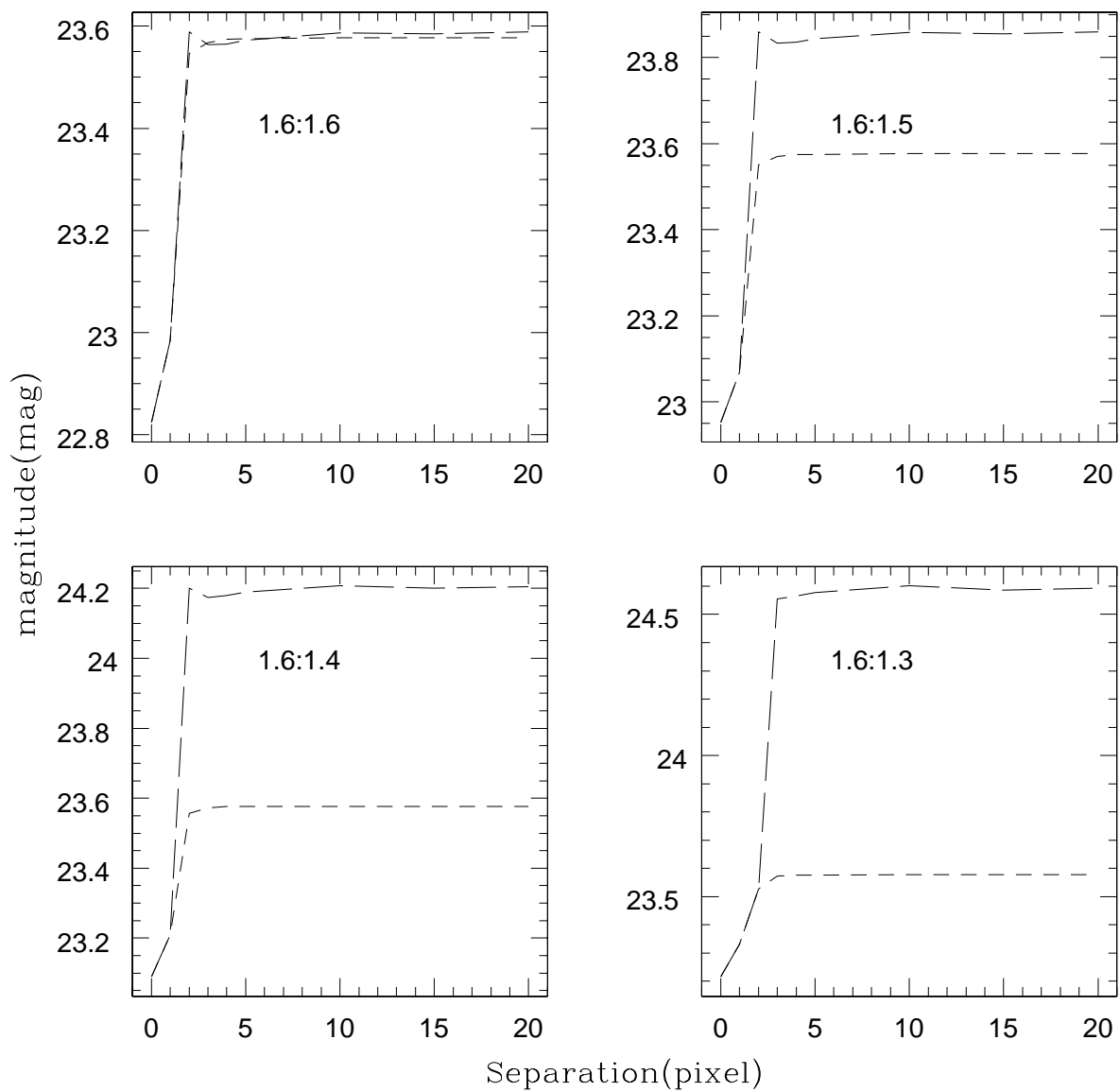}
\caption{Magnitude as a function of separation between the two
artificial stars. The label in each panel indicates the masses of the
input artificial stars. The short- and long-dashed lines represent
their recovered magnitudes. If the two stars are so close that only
one is recovered, we assign both stars the same magnitude. For equal
masses (a mass ratio of unity), the two stars can both be recovered if
their separation is greater than 2 pixels.}
\end{figure*}

\clearpage
\begin{figure*}
\includegraphics[width=\columnwidth]{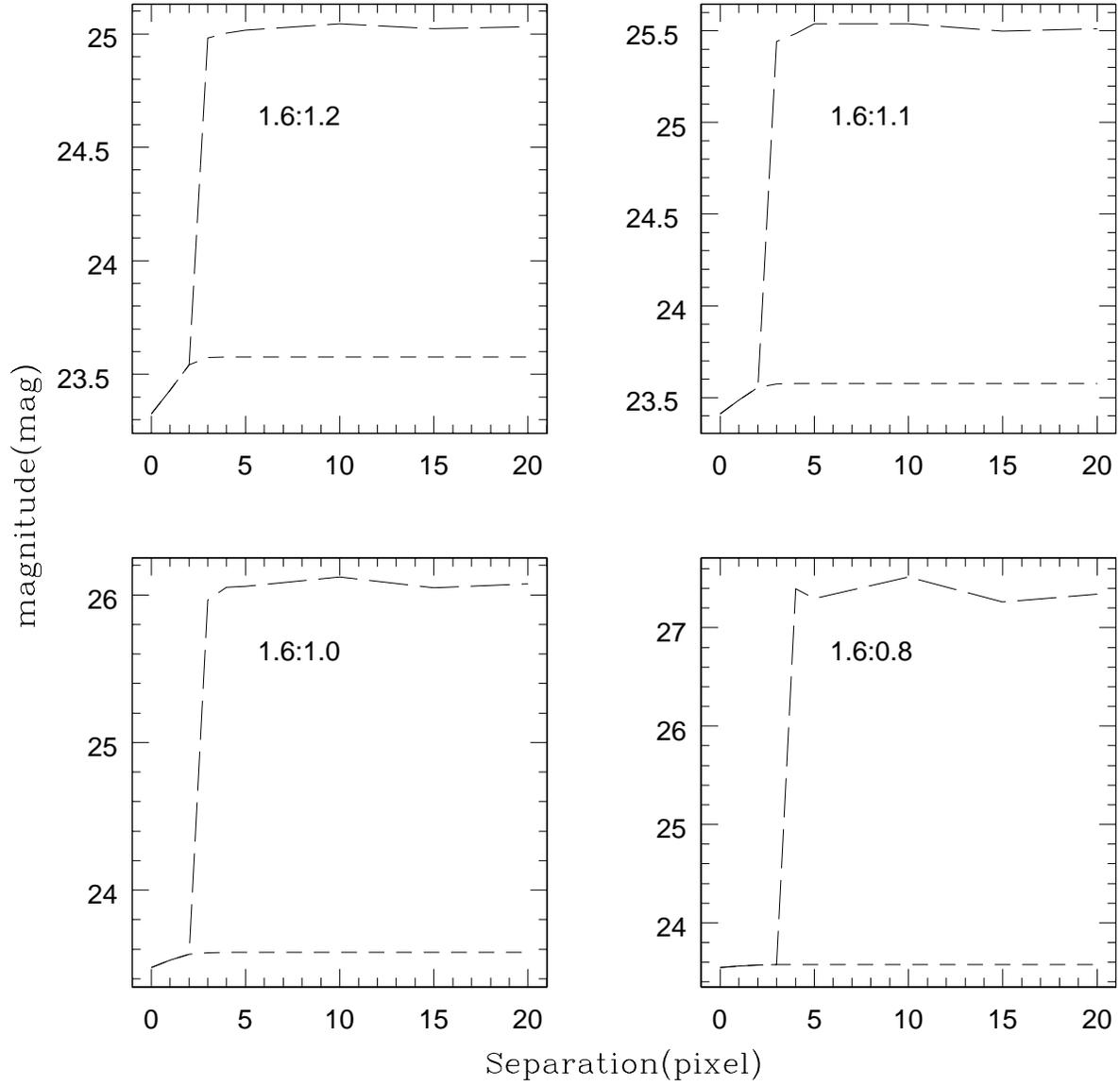}
\caption{Same as Fig. 7, but for different mass
ratios.}
\end{figure*}

\clearpage
\begin{figure*}
\includegraphics[width=\columnwidth]{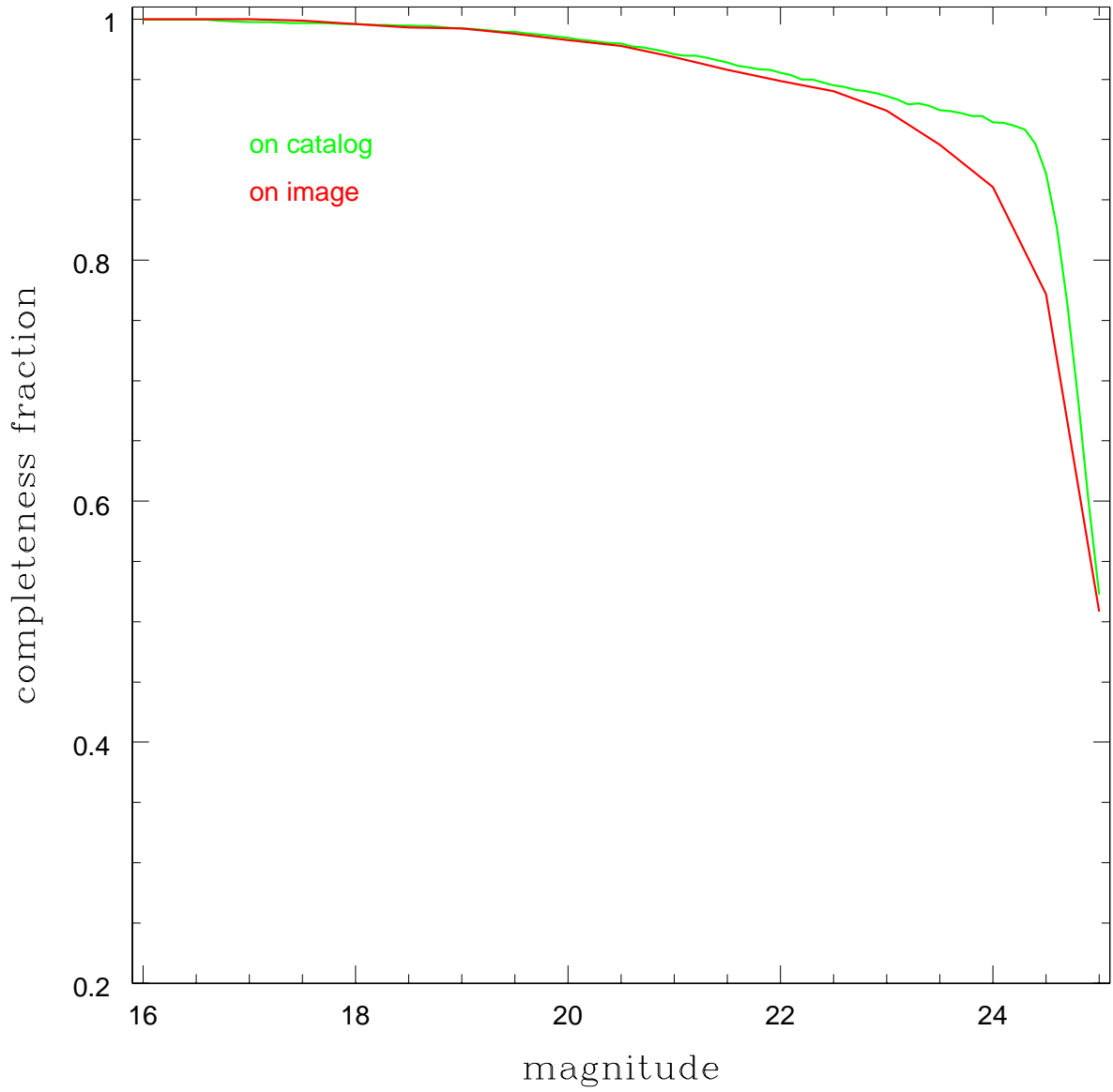}
\caption{Completeness fractions obtained from the two methods. }
\end{figure*}

\clearpage
\begin{figure*}
\includegraphics[width=\columnwidth]{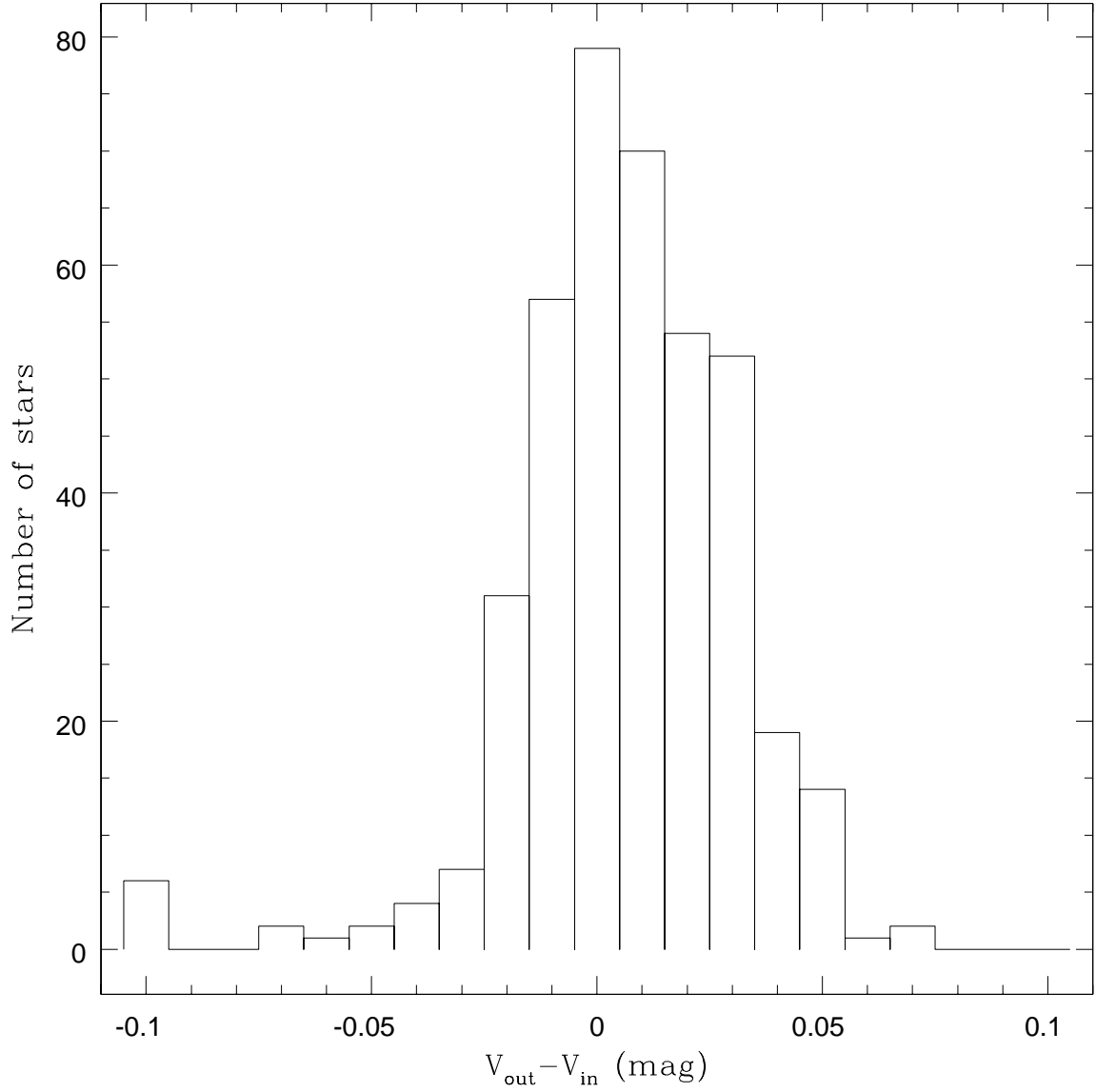}
\caption{Distribution of output $V$-band magnitudes of 500 recovered
stars using standard data-reduction techniques.}
\end{figure*}

\end{document}